\documentclass[11pt,draftcls,onecolumn]{IEEEtran}
\IEEEoverridecommandlockouts                              % This command is only
\usepackage{epsfig}
\usepackage{amsmath}
\usepackage{amssymb}
\usepackage{indentfirst}
\usepackage{cite}
\usepackage{bm}
\usepackage{graphicx}           % for including graphics
\usepackage{psfrag}             % for changing text in eps graphics
\usepackage{subfigure}          % for subfigures
\usepackage{empheq}
\usepackage{enumerate}

\usepackage{amsthm}

\usepackage{tikz}
\usetikzlibrary{arrows}

\newenvironment{proof}[1][Proof]%
  {\smallskip\par\noindent\textbf{#1\,:\ }}%
  {\hspace*{\fill} \rule{6pt}{6pt}\smallskip}
\newenvironment{proof*}[1][Proof]%
  {\medskip\par\noindent\textbf{#1\,:\ }}%
\usepackage{color}
\definecolor{gray}{RGB}{128,128,128}

\usepackage{hyperref}

\title{Multi-Agent Coordination for Distributed Transmit Beamforming}

\author{Jemin George, Anjaly Parayil and He Bai% <-this % stops a space
\thanks{Jemin~George and Anjaly Parayil are with the U.S. Army Research Laboratory, Adelphi, MD 20783, USA.
{\tt\small jemin.george.civ@mail.mil}}%
\thanks{He Bai is with the Department of Mechanical and Aerospace Engineering, Oklahoma State University, Stillwater, OK 74078 USA.
{\tt\small he.bai@okstate.edu}}%
}

\allowdisplaybreaks[4]

\begin{document}

\maketitle
\thispagestyle{empty}
\pagestyle{empty}

%%%%%%%%%%%%%%%%%%%%%%%%%%%%%%%%%%%%%%%%%%%%%%%%%%%%%%%%%%%%%%%%%%%%%%%%%%%%%%%%
\begin{abstract}
This paper presents the formulation and analysis of a two time-scale optimization algorithm for multi-agent coordination for the purpose of distributed beamforming. Each agent is assumed to be randomly positioned with respect to each other with random phase offsets and amplitudes. Agents are tasked with coordinate among themselves to position themselves and adjust their phase offset and amplitude such that they can construct a desired directed beam. Here we propose a two time-scale optimization algorithm that consists of a fast time-scale algorithm to solve for the amplitude and phase while a slow time-scale algorithm to solve for the control required to re-position the agents. The numerical results given here indicate that the proposed two time-scale approach is able to reconstruct a desired beam pattern.
\end{abstract}
%%%%%%%%%%%%%%%%%%%%%%%%%%%%%%%%%%%%%%%%%%%%%%%%%%%%%%%%%%%%%%%%%%%%%%%%%%%%%%%%

%~~~~~~~~~~~~~~~~~~~~~~~~~~~~~~~~~~~~~~~~~~~~~~~~~~~~~~~~~~~~~~~~~~~~~~~~~~~~~~~~~~~~~~~~~~~~~~~~~~~~~~~~~~~~~~~~~~~~~~
\section{Introduction}\label{sec:intro}
%~~~~~~~~~~~~~~~~~~~~~~~~~~~~~~~~~~~~~~~~~~~~~~~~~~~~~~~~~~~~~~~~~~~~~~~~~~~~~~~~~~~~~~~~~~~~~~~~~~~~~~~~~~~~~~~~~~~~~~

Distributed beamforming is concerned with the problem of cooperative communication where randomly located independent nodes coordinate among themselves to form a virtual antenna array. Although numerous studies on distributed beamforming have been carried out for over a decade, it was initially considered impractical due to the high complexity involved in modeling the generated beam pattern and the hardly achievable requirements on positioning and synchronization. Recent research results demonstrating the efficacy of distributed beamforming as a suitable solution for 5G communication systems such as mm-wave communication and machine to machine communications has further ignited the interest in this research field. The concept of distributed beamforming was conceived in early 2000’s by two independent pieces of research under the names collaborative beamforming \cite{ Ochiai2005} and distributed beamforming \cite{Barriac2004}. While initial research on collaborative beamforming focused on the beampattern analysis and the random array theory while assuming perfect phase synchronization among the nodes, the research on distributed beamforming focused only on the feasibility of achieving synchronization among distributed nodes and did not consider the significance of the physical array geometry and the beampattern. Over the years, the lines that separated the collaborative beamforming and distributed beamforming significantly blurred such that both the terms are now interchangeable.

Similar to the conventional antenna array beamforming, the distributed beamforming provides improvement in the received signal-to-noise ratio (SNR) compared to a point-to-point transmission. With a fixed radiated power at each antenna element (node), an ideal distributed beamformer with $n$ collaborating nodes will result in $n^2$ fold increase in the received power at the destination \cite{Ochiai2005}. Conversely, received power can be reduced by an order of $\frac{1}{n^2}$ for a fixed received power threshold. Thus distributed beamforming based collaborative communication drastically decreases the transmit power requirements allowing the individual nodes to conserve crucial resources and battery life especially in applications where the network is deployed at places where it difficult to replace or recharge the power source. Distributed beamforming has also shown to helps alleviate the long-distance transmission limitation in circumstances where it is unsuitable to layout sink node and multi-hop transmission.

Though much of the distributed beamforming works simply focus on achieving a desired SNR at the receiver, sophisticated distributed array techniques such as null-forming has also shown to be achieved through distributed beamforming as a solution to the covert communication problem~\cite{ Kumar2014, Goguri2016}. However, null-forming is a formidable problem due to its sensitivity to small phase errors. Furthermore, since null-forming fundamentally relies on a node’s transmitted signal cancelling the signals from all other transmitters, the amplitude and phase of the transmitted signal at each node cannot be chosen independently of the amplitudes and phases of other nodes. Therefore distributed null-forming algorithms often assume that each transmitter knows every transmitter’s complex channel gain to the receiver, in other words, global channel state information at each of the transmitters (CSIT).

There exists a plethora of literature on the concept of distributed transmit beamforming. For example, reference \cite{ Mudumbai2009} reviews several results in architectures, algorithms, and working prototypes available almost a decade ago to address the changes of coordinating the sources for information sharing and timing synchronization and, most crucially, distributed carrier synchronization so that the transmissions combine constructively at the destination. In order to ensure phase coherence of the radio frequency signals from the different transmitters in the presence of unknown phase offsets between the transmitters and unknown channel gains from the transmitters to the receiver, in \cite{Mudumbai2010}, authors propose a distributed adaptation scheme, where each transmitter independently makes a small random adjustment to its phase at each iteration, while the receiver broadcasts a single bit of feedback, indicating whether the signal-to-noise ratio (SNR) improved or worsened after the current iteration. Reference \cite{Liu2010} investigates linear beamforming techniques in relay networks with multiple independent sources, destinations and relay(s), where the goal is to determine the beamforming matrix to minimize the sum transmit power at the relays while meeting signal-to-interference (SINR) requirements at the destinations. In \cite{ Brown2012} authors describes a receiver-coordinated distributed transmission protocol for the joint beamforming and nullforming problem, in which the receive nodes feedback periodic channel measurements to the transmit cluster and the transmit nodes use this feedback to generate optimal channel predictions and then calculate a time-varying transmit vector that minimizes the average total power at the protected receivers while satisfying an average power constraint at the intended receiver during distributed transmission. Similarly, \cite{Fan2017} proposes a fast baseband transmit beamforming algorithm for the distributed antennas with one-bit feedback control using the received signal strength (RSS) at the receiver. An adaptive minimum variance distortion-less response (MVDR) beamformer for nonuniform linear arrays with enhanced degrees of freedom is presented in \cite{Yu2015} to enforce a unit response at the direction of the desired signal and places nulls in the directions of the interferences. In \cite{Kumar2017} authors consider the distributed joint beamforming and nullforming problem where $N$ single antenna transmitters must broadcast a common message signal by forming beams towards each of the single antenna receivers, while simultaneously forming nulls at another set of receivers. After formulating the problem as an unconstrained optimization problem to minimize the mean square error between the achieved and desired modulating amplitudes at the receivers, authors propose a gradient descent algorithm that utilizes a common feedback message, broadcasted by each of the receivers to all transmitters, consisting of a single complex number representing the amplitude of the aggregate (total) baseband received signal in the previous iteration. While most of the above mentioned work only considers the phase coherence of the radio frequency signals at each of the transmitters as the control variable, more resent works~\cite{ Chatzipanagiotis2014, Farazi2016, Muralidharan2018} focuses on both the transmitter position as well the phase offset. Finally a comprehensive survey of various distributed beamforming research, as well as its classifications, inherent features, constraints, challenges and the lessons learned from the shortcomings of previous research are summarized in \cite{Jayaprakasam2017}.

In this paper we consider the problem of beam matching, as opposed to previous methods for distributed and mobile beamformers that wish to maximize SINR at the client while minimizing transmit power \cite{chatzipanagiotis2012controlling}.  The reasoning's behind our proposed objective are two-fold; first, the transmit power given the proposed devices and frequencies used are minimal when compared to the power used to maneuver the transmit nodes.  With that being said, minimizing transmit power is not a major performance objective.  Second, we wish to have more precision with the formed beam that would provide built in null-forming for covert missions.\cite{farazi2016simultaneous} proposes a phase adjustment protocol to maintain perfect nulls at desired locations while mobile nodes alter their positions to improve the received power at the client. This method relies on random perturbations to the phase displacement vector and use of the heavy ball method.

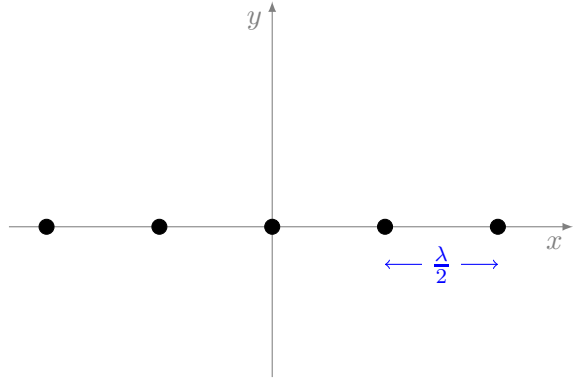
\begin{figure}[ht]
  \centering
  \begin{tikzpicture}
    \coordinate (Origin)   at (0,0);
    \coordinate (XAxisMin) at (-3.5,0);
    \coordinate (XAxisMax) at (4,0);
    \coordinate (YAxisMin) at (0,-2);
    \coordinate (YAxisMax) at (0,3);
    \draw [thin, gray,-latex] (XAxisMin) -- (XAxisMax) node [below left] {$x$};% Draw x axis
    \draw [thin, gray,-latex] (YAxisMin) -- (YAxisMax) node [below left] {$y$};;% Draw y axis

    %\clip (-5,-2) rectangle (5cm,4cm); % Clips the picture...

    \foreach \x in {-2,-1,...,2}{% Two indices running over each
      \foreach \y in {0}{% node on the grid we have drawn
        \node[draw,circle,inner sep=2pt,fill] at (1.5*\x,1.5*\y) {};
            % Places a dot at those points
      }
    }
    \draw[blue, <->] (1.5,-0.5) -- (3,-0.5) node [midway,fill=white] {$\frac{\lambda}{2}$};
  \end{tikzpicture}
  \caption{ESLA with $n=5$ and $d=\frac{\lambda}{2}$}
  \label{ESLA}
\end{figure}

Consider an equally spaced linear array (ESLA) of $n$ elements. See Fig.~\ref{ESLA}.
The general array factor can be written as
\begin{equation}\label{Ern:AF}
  AF(\theta) = \sum_{m=0}^{n-1} I_m\, e^{jk \mathbf{r}_m^\top \hat{\mathbf{r}}},
\end{equation}
where $\mathbf{r}_m$ is the vector to the $m$-th element, $\hat{\mathbf{r}}$ is a unit vector pointing in the direction of interest i.e.,
$$\hat{\mathbf{r}} = \begin{bmatrix}
                       \cos(\theta) & \sin(\theta)
                     \end{bmatrix}^\top,$$
$k$ is the wave number and $I_m$ is the element excitation with amplitude $a_m$ and linear phase gradient of $\alpha$ across the array, i.e.,
\begin{equation}\label{Ern:Im}
  I_m = a_m\, e^{jm\alpha}.
\end{equation}
For the ESLA given in Fig.~\ref{ESLA}, we have
$$\mathbf{r}_m = \begin{bmatrix}
                       (m-2)\frac{\lambda}{2} & 0
                     \end{bmatrix}^\top,$$
and
\begin{equation}\label{Ern:AF1}
  AF(\theta) = \sum_{m=0}^{n-1} a_m\, e^{j\left(m\alpha+k (m-2)\frac{\lambda}{2}\cos(\theta) \right)}.
\end{equation}
%%%%%%%%%%%%%%%%%%%%%%%%%%%%%%%%%%%%%%%%%%%%%%%%%%%%%%%%%%%%%%%%%%%%%%%%%%%%
\begin{figure}[!ht]
  \begin{centering}
      \includegraphics[width=.5\textwidth]{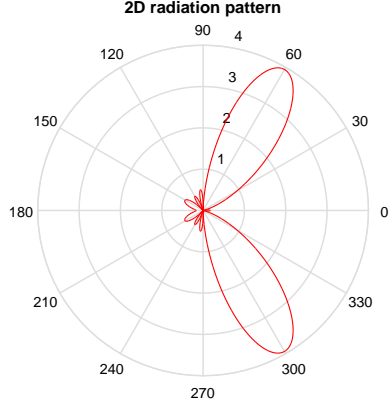}
      \caption{2-D Beam Pattern}
  \end{centering}\label{BeamPattern}
\end{figure}
%%%%%%%%%%%%%%%%%%%%%%%%%%%%%%%%%%%%%%%%%%%%%%%%%%%%%%%%%%%%%%%%%%%%%%%%%%%%
Figure \ref{BeamPattern} shows the beam pattern obtained for the 5 element ESLA with Binomial amplitude tapering and a phase gradient of $\alpha = -\pi/2$ at $40$ MHz.

The generalized array factor for $n$-elements located at $\begin{bmatrix} x_1 \\ y_1  \end{bmatrix}$ $\ldots$ $\begin{bmatrix} x_m \\ y_m  \end{bmatrix}$ $\ldots$ $\begin{bmatrix} x_n \\ y_n  \end{bmatrix}$ can be written as
\begin{equation}\label{Ern:AF0}
  AF(\theta) = \sum_{m=0}^{n-1} a_m\, e^{j\left(m\alpha+k x_m\cos(\theta) + k y_m \sin(\theta) \right)}.
\end{equation}
Assume there exists a receiver at location $\mathbf{p}\in\mathbb{R}^2$. The channel between the $m$-th element and the receiver is modeled as
\begin{equation}\label{Ern:Ch}
  c_m(\mathbf{p}) = \gamma\left( \mathbf{r}_m \right) \beta(d_m) e^{j k d_m},
\end{equation}
where $d_m = \| \mathbf{p} - \mathbf{r}_m \|_2$ is the distance between the $m$-th element and the receiver, $\gamma\left( \cdot \right): \mathbb{R}^2\mapsto\mathbb{R}$ captures multipath fading and $\beta\left( \cdot \right): \mathbb{R}\mapsto\mathbb{R}$ denotes the path loss. Here we model the path loss as
\begin{equation}
   \beta(d_m) = d_m^{-\mu/2},
\end{equation}
where $\mu$ is the path loss exponent. For the frequency we are considering we model the multipath gain as a random variable $\gamma_m$. If $\mathbf{p} = \rho \hat{\mathbf{r}}$, then the beam pattern taking the channel into consideration is
\begin{align}\label{Ern:AF2}
\begin{split}
  AF&(\rho,\theta) = \sum_{m=0}^{n-1}  \frac{a_m\, \gamma_m}{\left(d_m\right)^{\mu/2}}\,e^{j\left(m\alpha+ k x_m\cos(\theta) + k y_m \sin(\theta) + k d_m\right)},
 \end{split}
\end{align}
where $d_m = \| \rho \hat{\mathbf{r}} - \mathbf{r}_m \|_2$. Note that the channel could affect both amplitude and phase.

\section{Problem Formulation}
%~~~~~~~~~~~~~~~~~~~~~~~~~~~~~~~~~~~~~~~~~~~~~~~~~~~~~~~~~~~~~~~~~~~~~~~~~~~~~~~~~~~~~~~~~~~~~~~~~~~~~~~~~~~~~~~~~~~~~~

Consider the following desired array pattern constructed by $n$ fictitious agents located at $\begin{bmatrix} \bar{x}_0 \\ \bar{y}_{0}\end{bmatrix}$, $\ldots$, $\begin{bmatrix} \bar{x}_{n-1} \\ \bar{y}_{n-1}\end{bmatrix}$:
\begin{align}\label{Ern:AFd}
\begin{split}
  AF_d&(\rho,\theta) = \sum_{m=0}^{n-1} \frac{\bar{a}_m\, \bar{\gamma}_m}{\left(\bar{d}_m\right)^{\bar{\mu}/2}}\,e^{j\left(m\bar{\alpha}+ k \bar{x}_m\cos(\theta) + k \bar{y}_m \sin(\theta) + k \bar{d}_m\right)},
 \end{split}
\end{align}
where
$$ \bar{d}_m = \left\| \begin{bmatrix} \bar{x}_m \\ \bar{y}_m \end{bmatrix} - \rho \begin{bmatrix} \cos(\theta) \\ \sin(\theta)\end{bmatrix}\right\|_2, $$
$\bar{a}_m$, $\bar{\mu}$, and $\bar{\alpha}$ are nominal system values (these could be unknown parameters). Now $s$ mobile agents (array elements) located at $\mathbf{r}_0(t_0)$ $\ldots$ $\mathbf{r}_{s-1}(t_0)$ would like to construct the desired pattern such that
\begin{align}\label{Ern:AFd}
\begin{split}
  J &= \frac{1}{2}\,\int_{0}^{2\pi} \int_{\rho_a}^{\rho_b} \, \left\| |AF_d(\rho,\theta)| - |AF(\rho,\theta)| \right\|_2^2 \, d\theta\, d\rho + \sum_{m=0}^{s-1} \, \int_{t_0}^{t_f} \left[ \mathbf{r}_m(t)-\mathbf{r}_m(t_0)\right]^\top S_m \left[ \mathbf{r}_m(t)-\mathbf{r}_m(t_0)\right] dt
 \end{split}
\end{align}
is minimized. Here $S_m$ is a positive definite matrix. The true array factor is given as
\begin{align}\label{Ern:AFa}
\begin{split}
  AF&(\rho,\theta) = \sum_{m=0}^{s-1} \frac{{a}_m\, {\gamma}_m}{\left(d_m(t_f)\right)^{{\mu}/2}}\,e^{j\left({\alpha}_m+ k x_m(t_f)\cos(\theta) + k y_m(t_f) \sin(\theta) + k d_m(t_f)\right)},
 \end{split}
\end{align}
where $ \mathbf{r}_m(t_f) = \begin{bmatrix} x_m(t_f) & y_m(t_f) \end{bmatrix} ^\top$ is the actual element locations, and
$$ d_m(t_f) = \left\| \begin{bmatrix} {x}_m(t_f) \\ {y}_m(t_f) \end{bmatrix} - \rho \begin{bmatrix} \cos(\theta) \\ \sin(\theta)\end{bmatrix}\right\|_2.$$
Design parameters are $\alpha_0,\ldots,\alpha_{s-1}$, $a_0$, $\ldots$, $a_{s-1}$, $\mathbf{r}_0(t_f)$, $\ldots$, $\mathbf{r}_{s-1}(t_f)$.

\subsection{Simplification}

In practice, it is usually not required to match the beam pattern over a continuous space, but to rather match the beam at desired instances of $\rho$ and $\theta$ deemed valuable for either transmission or null forming. We start with discretizing the polar coordinates as $\{\theta_1,\ldots,\theta_l\}$ and $\{\rho_1,\ldots,\rho_{\ell}\}$. Thus the magnitude of the desired total array factor for all $(\rho_i, \theta_i)$ pairs, where $\theta_i \in \{\theta_1,\ldots,\theta_l\}$ and $\rho_i\in\{\rho_1,\ldots,\rho_{\ell}\}$, can be denoted as $f(\rho_i, \theta_i)$. For example, Fig.~\ref{BeamPattern} could present one such pattern of $f(\rho_i, \theta_i)$.

Now the resulting problem can be viewed as the following optimization problem
\begin{align}
\begin{split}\label{OptCont}
    &\min_{\bm{\alpha},\mathbf{a},\mathbf{r}(t_f)}\quad J =\frac{1}{2}\,\sum_{i} \| f(\rho_i, \theta_i) -  |AF(\rho_i, \theta_i, \bm{\alpha},\mathbf{a}, \mathbf{r}_m(t_f))| \|_2^2\\
     &\qquad \qquad\qquad \qquad + \sum_{m=0}^{s-1} \, \int_{t_0}^{t_f} \left[ \mathbf{r}_m(t)-\mathbf{r}_m(t_0)\right]^\top S_m \left[ \mathbf{r}_m(t)-\mathbf{r}_m(t_0)\right]
      \,\,dt.
    \end{split}
\end{align}

%~~~~~~~~~~~~~~~~~~~~~~~~~~~~~~~~~~~~~~~~~~~~~~~~~~~~~~~~~~~~~~~~~~~~~~~~~~~~~~~~~~~~~~~~~~~~~~~~~~~~~~~~~~~~~~~~~~~~~~
\section{A proposed solution}
%~~~~~~~~~~~~~~~~~~~~~~~~~~~~~~~~~~~~~~~~~~~~~~~~~~~~~~~~~~~~~~~~~~~~~~~~~~~~~~~~~~~~~~~~~~~~~~~~~~~~~~~~~~~~~~~~~~~~~~

The proposed solution consists of a two time-scale process. A fast time-scale optimization process to identify the amplitude and phase, while a slow time-scale process to relocate the agents if needed.
%Here we propose a two step solution: the first step involves an optimization with stationary (current) locations to find the beamforming weights such that $\sum_{i} \| f(\rho_i, \theta_i) -  |AF(\rho_i, \theta_i, \bm{\alpha},\mathbf{a}, \mathbf{r}_m(t))| \|_2$ is minimized. If the agents are not able to obtain the desired pattern by adjusting the weights, then go to step two. The second step consist of a feedback solutions to move the agents.

Before we proceed we simplify the problem by assuming that the number of antenna-elements involved in the construction of desired beam pattern is same as the number of mobile-agents involved in distributed beamforming, i.e., $n = s$.

%~~~~~~~~~~~~~~~~~~~~~~~~~~~~~~~~~~~~~~~~~~~~~~~~~~~~~~~~~~~~~~~~~~~~~~~~~~~~~~~~~~~~~~~~~~~~~~~~~~~~~~~~~~~~~~~~~~~~~~
\subsection{Fast-Scale Optimization}
%~~~~~~~~~~~~~~~~~~~~~~~~~~~~~~~~~~~~~~~~~~~~~~~~~~~~~~~~~~~~~~~~~~~~~~~~~~~~~~~~~~~~~~~~~~~~~~~~~~~~~~~~~~~~~~~~~~~~~~

Given the current location of the agents, $\mathbf{r}_0(t)$ $\ldots$ $\mathbf{r}_{s-1}(t)$, $AF(\rho_i,\theta_i)$ can be written as

\begin{align}\label{Ern:AFa2}
\begin{split}
  AF&(\rho_i,\theta_i) = \sum_{m=0}^{s-1} \frac{ {a_m\,\gamma}_m}{\left(d_{m_i}(t)\right)^{{\mu}/2}}\,e^{j\left( \alpha_m + k x_m(t)\cos(\theta_i) + k y_m(t) \sin(\theta_i) + k d_{m_i}(t)\right)},
 \end{split}
\end{align}
where
\begin{align}
    d_{m_i}(t) = \left\| \begin{bmatrix} {x}_m(t) \\ {y}_m(t) \end{bmatrix} - \rho_i \begin{bmatrix} \cos(\theta_i) \\ \sin(\theta_i)\end{bmatrix}\right\|_2.
\end{align}
Thus an optimization problem can be posed as
\begin{align}\label{Eqn:Opt1}
\begin{split}
 \min_{\bm{\alpha},\mathbf{a}}\,\,
\frac{1}{2}\, \sum_{i} \left\| f(\rho_i, \theta_i) -  \left| \sum_{m=0}^{s-1} \frac{ {a_m\,\gamma}_m}{\left(d_{m_i}(t)\right)^{{\mu}/2}}\,e^{j\left( \alpha_m + k x_m(t)\cos(\theta_i) + k y_m(t) \sin(\theta_i) + k d_{m_i}(t)\right)} \right| \right\|_2^2.
 \end{split}
\end{align}

Let
$$ \Phi_i\left(\bm{\alpha},\mathbf{a},\rho_i,\theta_i,t\right) = \frac{1}{2}\, \left\| f(\rho_i, \theta_i) -  \left| \sum_{m=0}^{s-1} \frac{ {a_m\,\gamma}_m}{\left(d_{m_i}(t)\right)^{{\mu}/2}}\,e^{j\left( \alpha_m + k x_m(t)\cos(\theta_i) + k y_m(t) \sin(\theta_i) + k d_{m_i}(t)\right)} \right| \right\|_2^2$$
Now the above optimization problem can be rewritten as
\begin{align}\label{Eqn:Opt2}
\begin{split}
 \min_{\bm{\alpha},\mathbf{a}}\,\,
 \sum_{i} \Phi_i\left(\bm{\alpha},\mathbf{a},\rho_i,\theta_i,t\right)
 \end{split}
\end{align}
Note that the objective $\Phi_i$ is time-varying since it changes with agent location. Ideally, we would like to keep the agents stationary while solving for the optimal beamforming weights. However the agents are constantly moving and therefore we propose the following fast gradient flow to solve for the weights:
\begin{align}
\epsilon \dot{\mathbf{a}}(t) &=  - \sum_{i} \nabla_{\mathbf{a}} \Phi_i\left(\bm{\alpha},\mathbf{a},\rho_i,\theta_i,t\right), \epsilon \ll 1, \quad \textnormal{and}\\
 \epsilon \dot{\bm{\alpha}}(t) &=  - \sum_{i} \nabla_{\bm{\alpha}} \Phi_i\left(\bm{\alpha},\mathbf{a},\rho_i,\theta_i,t\right)
\end{align}
In other words, we assume that the agents are moving sufficiently slow comparing with the above optimization algorithm, i.e., $\dot r_m \sim O(\epsilon)$.
The gradients $\nabla_{\mathbf{a}} \Phi_i\left(\bm{\alpha},\mathbf{a},\rho_i,\theta_i,t\right)$ are calculated as
\begin{align}
\begin{split}
    \nabla_{\mathbf{a}} \Phi_i\left(\bm{\alpha},\mathbf{a},\rho_i,\theta_i,t\right) &= \left(\left| \sum_{m=0}^{s-1} \frac{ {a_m\,\gamma}_m}{\left(d_{m_i}(t)\right)^{{\mu}/2}}\,e^{j\left( \alpha_m + k x_m(t)\cos(\theta_i) + k y_m(t) \sin(\theta_i) + k d_{m_i}(t)\right)} \right|-f(\rho_i, \theta_i)\right) \\
    &\qquad \qquad \times
    \frac{\partial  \left| \displaystyle \sum_{m=0}^{s-1} \frac{ {a_m\,\gamma}_m}{\left(d_{m_i}(t)\right)^{{\mu}/2}}\,e^{j\left( \alpha_m + k x_m(t)\cos(\theta_i) + k y_m(t) \sin(\theta_i) + k d_{m_i}(t)\right)} \right|}{\partial \mathbf{a}}
\end{split}
\end{align}
Define
\begin{align}
    \zeta_{m_i}(t) =  k x_m(t)\cos(\theta_i) + k y_m(t) \sin(\theta_i) + k d_{m_i}(t).
\end{align}
Now using the Euler identity we have
\begin{align}
\begin{split}
   \left| \displaystyle \sum_{m=0}^{s-1} \frac{ {a_m\,\gamma}_m}{\left(d_{m_i}(t)\right)^{{\mu}/2}}\,e^{j\left( \alpha_m + \zeta_{m_i}(t) \right)} \right| &= \Bigg( \left( \sum_{m=0}^{s-1} \frac{ {a_m\gamma}_m}{\left(d_{m_i}(t)\right)^{{\mu}/2}}\, \cos\left(\alpha_m +\zeta_{m_i}(t)\right) \right)^2  \\& \qquad + \left( \sum_{m=0}^{s-1} \frac{ {a_m\gamma}_m}{\left(d_{m_i}(t)\right)^{{\mu}/2}}\, \sin\left(\alpha_m +\zeta_{m_i}(t)\right)\right)^2 \Bigg)^{1/2}
\end{split}
\end{align}
Let
\begin{align}
   \bm{u}_i(t) = \begin{bmatrix} \frac{ {\gamma}_0}{\left(d_{0_i}(t)\right)^{{\mu}/2}}\, \cos\left(\alpha_0 +\zeta_{0_i}(t)\right) \\
   \vdots\\
    \frac{ {\gamma}_{s-1}}{\left(d_{{s-1}_i}(t)\right)^{{\mu}/2}}\, \cos\left(\alpha_{s-1} +\zeta_{{s-1}_i}(t)\right)
   \end{bmatrix}
   \quad \textnormal{and} \quad
   \bm{v}_i(t) = \begin{bmatrix} \frac{ {\gamma}_0}{\left(d_{0_i}(t)\right)^{{\mu}/2}}\, \sin\left(\alpha_{0} +\zeta_{0_i}(t)\right) \\
   \vdots\\
    \frac{ {\gamma}_{s-1}}{\left(d_{{s-1}_i}(t)\right)^{{\mu}/2}}\, \sin\left(\alpha_{s-1} +\zeta_{{s-1}_i}(t)\right)
   \end{bmatrix}
\end{align}
Thus we have
\begin{align}
   \left| \displaystyle \sum_{m=0}^{s-1} \frac{ {a_m\,\gamma}_m}{\left(d_{m_i}(t)\right)^{{\mu}/2}}\,e^{j\left( \alpha_m + \zeta_{m_i}(t) \right)} \right| &= \left(  \mathbf{a}^\top\bm{u}_i(t)\bm{u}_i^\top(t)\mathbf{a}  + \mathbf{a}^\top\bm{v}_i(t)\bm{v}_i^\top(t)\mathbf{a} \right)^{1/2}
\end{align}
and
\begin{align}
\begin{split}
   \frac{\partial  \left| \displaystyle \sum_{m=0}^{s-1} \frac{ {a_m\,\gamma}_m}{\left(d_{m_i}(t)\right)^{{\mu}/2}}\,e^{j\left( \alpha_m + \zeta_{m_i}(t) \right)} \right| }{\partial \mathbf{a}} &= \frac{\mathbf{a}^\top\bm{u}_i(t)}{\sqrt{\mathbf{a}^\top\bm{u}_i(t)\bm{u}_i^\top(t)\mathbf{a}  + \mathbf{a}^\top\bm{v}_i(t)\bm{v}_i^\top(t)\mathbf{a}}}\bm{u}_i(t) \\
   &\qquad + \frac{\mathbf{a}^\top\bm{v}_i(t)}{\sqrt{\mathbf{a}^\top\bm{u}_i(t)\bm{u}_i^\top(t)\mathbf{a}  + \mathbf{a}^\top\bm{v}_i(t)\bm{v}_i^\top(t)\mathbf{a}}}\bm{v}_i(t).
  \end{split}
\end{align}
Therefore
\begin{align}
\begin{split}
    \nabla_{\mathbf{a}} \Phi_i\left(\bm{\alpha},\mathbf{a},\rho_i,\theta_i,t\right) &= \frac{\left( \left| \displaystyle \sum_{m=0}^{s-1} \frac{ {a_m\,\gamma}_m}{\left(d_{m_i}(t)\right)^{{\mu}/2}}\,e^{j\left( \alpha_m + \zeta_{m_i}(t) \right)} \right|-f(\rho_i, \theta_i)\right)\mathbf{a}^\top\bm{u}_i(t)}{\left| \displaystyle \sum_{m=0}^{s-1} \frac{ {a_m\,\gamma}_m}{\left(d_{m_i}(t)\right)^{{\mu}/2}}\,e^{j\left( \alpha_m + \zeta_{m_i}(t) \right)} \right|}\bm{u}_i(t) \\
   &\qquad + \frac{\left( \left| \displaystyle \sum_{m=0}^{s-1} \frac{ {a_m\,\gamma}_m}{\left(d_{m_i}(t)\right)^{{\mu}/2}}\,e^{j\left( \alpha_m + \zeta_{m_i}(t) \right)} \right|-f(\rho_i, \theta_i)\right)\mathbf{a}^\top\bm{v}_i(t)}{\left| \displaystyle \sum_{m=0}^{s-1} \frac{ {a_m\,\gamma}_m}{\left(d_{m_i}(t)\right)^{{\mu}/2}}\,e^{j\left( \alpha_m + \zeta_{m_i}(t) \right)} \right|}\bm{v}_i(t).
\end{split}
\end{align}
Similarly the gradients $\nabla_{\bm{\alpha}} \Phi_i\left(\bm{\alpha},\mathbf{a},\rho_i,\theta_i,t\right)$ are calculated as
\begin{align}
\begin{split}
    \nabla_{\bm{\alpha}} \Phi_i\left(\bm{\alpha},\mathbf{a},\rho_i,\theta_i,t\right) &= \left( \left| \displaystyle \sum_{m=0}^{s-1} \frac{ {a_m\,\gamma}_m}{\left(d_{m_i}(t)\right)^{{\mu}/2}}\,e^{j\left( \alpha_m + \zeta_{m_i}(t) \right)} \right|-f(\rho_i, \theta_i)\right) \frac{\partial  \left| \displaystyle \sum_{m=0}^{s-1} \frac{ {a_m\,\gamma}_m}{\left(d_{m_i}(t)\right)^{{\mu}/2}}\,e^{j\left( \alpha_m + \zeta_{m_i}(t) \right)} \right| }{\partial \bm{\alpha}}
\end{split}
\end{align}
Note
\begin{align}
\begin{split}
   \frac{\partial  \left| \displaystyle \sum_{m=0}^{s-1} \frac{ {a_m\,\gamma}_m}{\left(d_{m_i}(t)\right)^{{\mu}/2}}\,e^{j\left( \alpha_m + \zeta_{m_i}(t) \right)} \right| }{\partial \bm{\alpha}} &= \frac{\mathbf{a}^\top\bm{u}_i(t)}{\sqrt{\mathbf{a}^\top\bm{u}_i(t)\bm{u}_i^\top(t)\mathbf{a}  + \mathbf{a}^\top\bm{v}_i(t)\bm{v}_i^\top(t)\mathbf{a}}}\frac{\partial\mathbf{a}^\top\bm{u}_i(t)}{\partial \bm{\alpha}} \\
   &\qquad + \frac{\mathbf{a}^\top\bm{v}_i(t)}{\sqrt{\mathbf{a}^\top\bm{u}_i(t)\bm{u}_i^\top(t)\mathbf{a}  + \mathbf{a}^\top\bm{v}_i(t)\bm{v}_i^\top(t)\mathbf{a}}}\frac{\partial\mathbf{a}^\top\bm{v}_i(t)}{\partial \bm{\alpha}}
  \end{split}
\end{align}
where
\begin{align}
\frac{\partial\mathbf{a}^\top\bm{u}_i(t)}{\partial \bm{\alpha}} = -\left(\mathbf{a} \circ \bm{v}_i(t)\right) \quad \textnormal{and}\quad
\frac{\partial\mathbf{a}^\top\bm{v}_i(t)}{\partial \bm{\alpha}} = \left(\mathbf{a} \circ \bm{u}_i(t)\right).
\end{align}
Here $\circ$ denotes the Hadamard product or the element-wise product. Therefore
\begin{align}
\begin{split}
    \nabla_{\bm f{\alpha}} \Phi_i\left(\bm{\alpha},\mathbf{a},\rho_i,\theta_i,t\right) &= -\frac{\left( \left| \displaystyle \sum_{m=0}^{s-1} \frac{ {a_m\,\gamma}_m}{\left(d_{m_i}(t)\right)^{{\mu}/2}}\,e^{j\left( \alpha_m + \zeta_{m_i}(t) \right)} \right|-f(\rho_i, \theta_i)\right)\mathbf{a}^\top\bm{u}_i(t)}{\left| \displaystyle \sum_{m=0}^{s-1} \frac{ {a_m\,\gamma}_m}{\left(d_{m_i}(t)\right)^{{\mu}/2}}\,e^{j\left( \alpha_m + \zeta_{m_i}(t) \right)} \right|}\left(\mathbf{a} \circ \bm{v}_i(t)\right) \\
   &\qquad + \frac{\left( \left| \displaystyle \sum_{m=0}^{s-1} \frac{ {a_m\,\gamma}_m}{\left(d_{m_i}(t)\right)^{{\mu}/2}}\,e^{j\left( \alpha_m + \zeta_{m_i}(t) \right)} \right|-f(\rho_i, \theta_i)\right)\mathbf{a}^\top\bm{v}_i(t)}{\left| \displaystyle \sum_{m=0}^{s-1} \frac{ {a_m\,\gamma}_m}{\left(d_{m_i}(t)\right)^{{\mu}/2}}\,e^{j\left( \alpha_m + \zeta_{m_i}(t) \right)} \right|}\left(\mathbf{a} \circ \bm{u}_i(t)\right).
\end{split}
\end{align}

\subsection{Slow-Scale Motion Planning}

We rewrite the slow-scale optimization in \eqref{OptCont} as
\begin{align}\label{OptCont1}
\begin{split}
    &\min_{\mathbf{r}(t)}\quad J =  \sum_{i} \Phi_i\left(\bm{\alpha},\mathbf{a},\rho_i,\theta_i,t\right) + \sum_{m=1}^{s} \, \int_{t_0}^{t} \left[ \mathbf{r}_m(\tau)-\mathbf{r}_m(t_0)\right]^\top S_m \left[ \mathbf{r}_m(\tau)-\mathbf{r}_m(t_0)\right]  \,\,d\tau.
    \end{split}
\end{align}
Thus we propose the following gradient flow to solve for the control:
\begin{align}
 \dot{\mathbf{r}}_m(t) =  - \nabla_{\mathbf{r}_m} J,\quad \forall m=0,\ldots,s-1
\end{align}
Substituting \eqref{OptCont1}, the above equation can be rewritten as
\begin{align}
 \dot{\mathbf{r}}_m(t) &=  - \sum_{i} \nabla_{\mathbf{r}_m} \Phi_i\left(\bm{\alpha},\mathbf{a},\rho_i,\theta_i,t\right) + \mathbf{v}_m(t) \\
 \dot{\mathbf{v}}_m(t) &= -2S_m \left( \mathbf{r}_m(t)-\mathbf{r}_m(t_0)\right).
\end{align}
The gradients $\nabla_{\mathbf{r}_m} \Phi_i\left(\bm{\alpha},\mathbf{a},\rho_i,\theta_i,t\right)$ are calculated as
\begin{align}
    \nabla_{\mathbf{r}_m} \Phi_i\left(\bm{\alpha},\mathbf{a},\rho_i,\theta_i,t\right)&= \left( \left| \displaystyle \sum_{m=0}^{s-1} \frac{ {a_m\,\gamma}_m}{\left(d_{m_i}(t)\right)^{{\mu}/2}}\,e^{j\left( \alpha_m + \zeta_{m_i}(t) \right)} \right|-f(\rho_i, \theta_i)\right)\frac{\partial  \left| \displaystyle \sum_{m=0}^{s-1} \frac{ {a_m\,\gamma}_m}{\left(d_{m_i}(t)\right)^{{\mu}/2}}\,e^{j\left( \alpha_m + \zeta_{m_i}(t) \right)} \right|}{\partial \mathbf{r}_m}
\end{align}
Note
\begin{align}
\begin{split}
   \frac{\partial  \left| \displaystyle \sum_{m=0}^{s-1} \frac{ {a_m\,\gamma}_m}{\left(d_{m_i}(t)\right)^{{\mu}/2}}\,e^{j\left( \alpha_m + \zeta_{m_i}(t) \right)} \right|}{\partial \mathbf{r}_m} &= \frac{\mathbf{a}^\top\bm{u}_i(t)}{\sqrt{\bm{u}_i^\top(t)\mathbf{a}\mathbf{a}^\top\bm{u}_i(t)  + \bm{v}_i^\top(t)\mathbf{a}\mathbf{a}^\top\bm{v}_i(t)}}\frac{\partial \mathbf{a}^\top\bm{u}_i(t)}{\partial \mathbf{r}_m} \\
   &\qquad + \frac{\mathbf{a}^\top\bm{v}_i(t)}{\sqrt{\bm{u}_i^\top(t)\mathbf{a}\mathbf{a}^\top\bm{u}_i(t)  + \bm{v}_i^\top(t)\mathbf{a}\mathbf{a}^\top\bm{v}_i(t)}}\frac{\partial \mathbf{a}^\top\bm{v}_i(t)}{\partial \mathbf{r}_m}.
  \end{split}
\end{align}
Also note
\begin{align}
\begin{split}
   \frac{\partial \mathbf{a}^\top\bm{u}_i(t)}{\partial \mathbf{r}_m} =
   \begin{bmatrix}
   \displaystyle\frac{\partial  a_m {\gamma}_m\left(d_{m_i}(t)\right)^{{-\mu}/2}\, \cos\left( \alpha_m + \zeta_{m_i}(t) \right)}{\partial x_m(t)}\\
   \displaystyle\frac{\partial  a_m {\gamma}_m\left(d_{m_i}(t)\right)^{{-\mu}/2}\, \cos\left( \alpha_m + \zeta_{m_i}(t) \right)}{\partial y_m(t)}
   \end{bmatrix}
  \end{split}
\end{align}
and
\begin{align}
\begin{split}
   \frac{\partial \mathbf{a}^\top\bm{v}_i(t)}{\partial \mathbf{r}_m} =
   \begin{bmatrix}
   \displaystyle\frac{\partial  a_m {\gamma}_m\left(d_{m_i}(t)\right)^{{-\mu}/2}\, \sin\left( \alpha_m + \zeta_{m_i}(t) \right)}{\partial x_m(t)}\\
   \displaystyle\frac{\partial  a_m {\gamma}_m\left(d_{m_i}(t)\right)^{{-\mu}/2}\, \sin\left( \alpha_m + \zeta_{m_i}(t) \right)}{\partial y_m(t)}
   \end{bmatrix}
  \end{split}
\end{align}
Recall
$$d_{m_i}(t) = \left\| \begin{bmatrix} {x}_m(t) \\ {y}_m(t) \end{bmatrix} - \rho_i \begin{bmatrix} \cos(\theta_i) \\ \sin(\theta_i)\end{bmatrix}\right\|_2$$
and
$$\zeta_{m_i}(t) =  k x_m(t)\cos(\theta_i) + k y_m(t) \sin(\theta_i) + k d_{m_i}(t).$$
Now the partials can be computed as
\begin{align}
    \begin{split}
        \displaystyle\frac{\partial  a_m {\gamma}_m\left(d_{m_i}(t)\right)^{{-\mu}/2}\, \cos\left( \alpha_m + \zeta_{m_i}(t) \right)}{\partial x_m(t)}
        = &-\displaystyle\frac{\mu\, a_m \gamma_m \cos\left( \alpha_m + \zeta_{m_i}(t) \right)\left( 2x_m(t) - 2\rho_i\cos(\theta_i)\right)}{4\left(d_{m_i}(t)\right)^{\mu/2+2}}\\
        &\, -\displaystyle\frac{a_m {\gamma}_m\sin\left( \alpha_m + \zeta_{m_i}(t) \right)\left(k\cos(\theta_i)+\frac{k\left(2x_m(t)-2\rho_i\cos(\theta_i)\right)}{2d_{m_i}(t)}\right)}{\left(d_{m_i}(t)\right)^{{\mu}/2}}
    \end{split}\\
    \begin{split}
        \displaystyle\frac{\partial  a_m {\gamma}_m\left(d_{m_i}(t)\right)^{{-\mu}/2}\, \cos\left( \alpha_m + \zeta_{m_i}(t) \right)}{\partial y_m(t)}
        = &-\displaystyle\frac{\mu\, a_m \gamma_m \cos\left( \alpha_m + \zeta_{m_i}(t) \right)\left( 2y_m(t) - 2\rho_i\sin(\theta_i)\right)}{4\left(d_{m_i}(t)\right)^{\mu/2+2}}\\
        &\,-\displaystyle\frac{a_m {\gamma}_m\sin\left( \alpha_m + \zeta_{m_i}(t) \right)\left(k\sin(\theta_i)+\frac{k\left(2y_m(t)-2\rho_i\sin(\theta_i)\right)}{2d_{m_i}(t)}\right)}{\left(d_{m_i}(t)\right)^{{\mu}/2}}
    \end{split}\\
    \begin{split}
        \displaystyle\frac{\partial  a_m {\gamma}_m\left(d_{m_i}(t)\right)^{{-\mu}/2}\, \sin\left( \alpha_m + \zeta_{m_i}(t) \right)}{\partial x_m(t)}
        = &-\displaystyle\frac{\mu\, a_m \gamma_m \sin\left( \alpha_m + \zeta_{m_i}(t) \right)\left( 2x_m(t) - 2\rho_i\cos(\theta_i)\right)}{4\left(d_{m_i}(t)\right)^{\mu/2+2}}\\
        &\,+\displaystyle\frac{a_m {\gamma}_m\cos\left( \alpha_m + \zeta_{m_i}(t) \right)\left(k\cos(\theta_i)+\frac{k\left(2x_m(t)-2\rho_i\cos(\theta_i)\right)}{2d_{m_i}(t)}\right)}{\left(d_{m_i}(t)\right)^{{\mu}/2}}
    \end{split}\\
    \begin{split}
        \displaystyle\frac{\partial  a_m {\gamma}_m\left(d_{m_i}(t)\right)^{{-\mu}/2}\, \sin\left( \alpha_m + \zeta_{m_i}(t) \right)}{\partial y_m(t)}
        = &-\displaystyle\frac{\mu\, a_m \gamma_m \sin\left( \alpha_m + \zeta_{m_i}(t) \right)\left( 2y_m(t) - 2\rho_i\sin(\theta_i)\right)}{4\left(d_{m_i}(t)\right)^{\mu/2+2}}\\
        &\,+\displaystyle\frac{a_m {\gamma}_m\cos\left( \alpha_m + \zeta_{m_i}(t) \right)\left(k\sin(\theta_i)+\frac{k\left(2y_m(t)-2\rho_i\sin(\theta_i)\right)}{2d_{m_i}(t)}\right)}{\left(d_{m_i}(t)\right)^{{\mu}/2}}
    \end{split}
\end{align}

\subsection{Summary}

In summary the proposed solution consists of the following process:

\begin{itemize}
    \item Select an appropriate $\epsilon \ll 1$
    \item Simultaneously solve the
    \begin{itemize}
        \item Fast gradient flow for amplitude and phase
            \begin{align}
                \epsilon \dot{\mathbf{a}}(t) &=  - \sum_{i} \nabla_{\mathbf{a}} \Phi_i\left(\bm{\alpha},\mathbf{a},\rho_i,\theta_i,t\right),  \quad \textnormal{and}\\
                 \epsilon \dot{\bm{\alpha}}(t) &=  - \sum_{i} \nabla_{\bm{\alpha}} \Phi_i\left(\bm{\alpha},\mathbf{a},\rho_i,\theta_i,t\right)
            \end{align}
        \item Slow gradient flow for control input
        \begin{align}
             \dot{\mathbf{r}}_m(t) &=  - \sum_{i} \nabla_{\mathbf{r}_m} \Phi_i\left(\bm{\alpha},\mathbf{a},\rho_i,\theta_i,t\right) + \mathbf{v}_m(t) \\
             \dot{\mathbf{v}}_m(t) &= -2S_m \left( \mathbf{r}_m(t)-\mathbf{r}_m(t_0)\right).
        \end{align}
    \end{itemize}
\end{itemize}

\section{Numerical Results}

\begin{figure}[!hb]
  \begin{centering}
      \subfigure[Desired beam pattern]{
      \includegraphics[width=.45\textwidth]{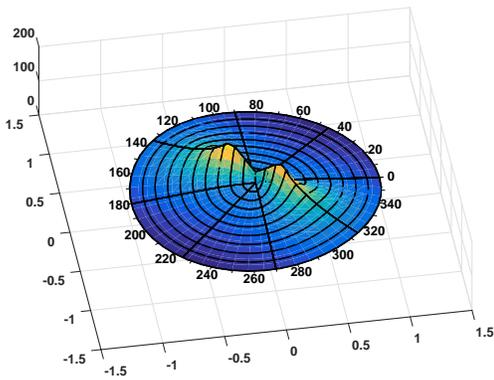}\label{Fig1}}
      \subfigure[Initial beam pattern]{
      \includegraphics[width=.45\textwidth]{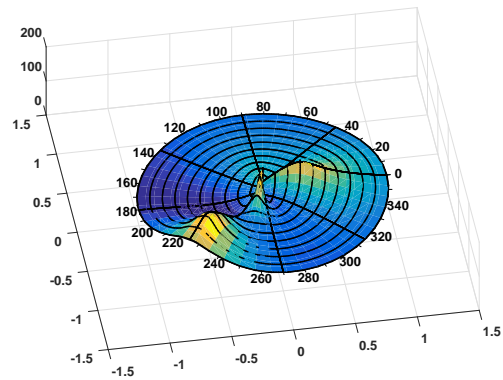}\label{Fig2}}
      \subfigure[Final constructed beam pattern]{
      \includegraphics[width=.55\textwidth]{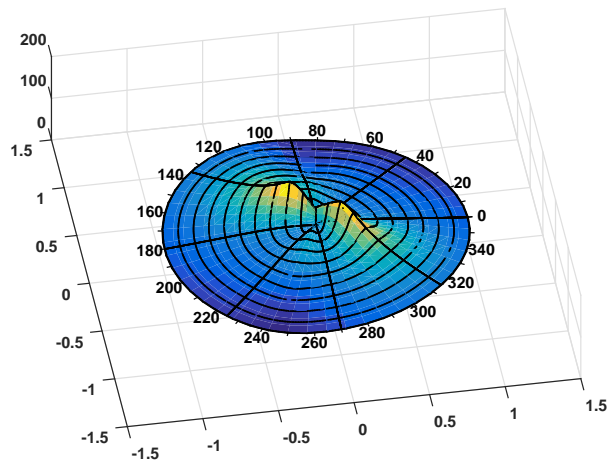}\label{Fig3}}
      \caption{2D beam pattern reconstructed using the proposed two time-scale approach}
      \label{Results1}
  \end{centering}
\end{figure}

For numerical simulations we consider a $5$ element beamforming array transmitting at 40 MHz. Given in Fig.~\ref{Fig1} is the desired beam pattern obtained by a linear array consisting of 5 elements. Initially the agents are randomly positioned and they have random phase offsets. The initial beam pattern obtained from the initial agent positions and phase offsets are given in Fig.~\ref{Fig2}. Finally given in Fig.~\ref{Fig3} is the reconstructed beam pattern obtained by implementing the proposed two time-scale optimization scheme. Note that the reconstructed beam pattern clearly matches the desired beam pattern.

%%%%%%%%%%%%%%%%%%%%%%%%%%%%%%%%%%%%%%%%%%%%%%%%%%%%%%%%%%%%%%%%%%%%%%%%%%%%
%%%%%%%%%%%%%%%%%%%%%%%%%%%%%%%%%%%%%%%%%%%%%%%%%%%%%%%%%%%%%%%%%%%%%%%%%%%%

\begin{figure}[!ht]
  \begin{centering}
      \subfigure[Amplitude]{
      \includegraphics[width=.45\textwidth]{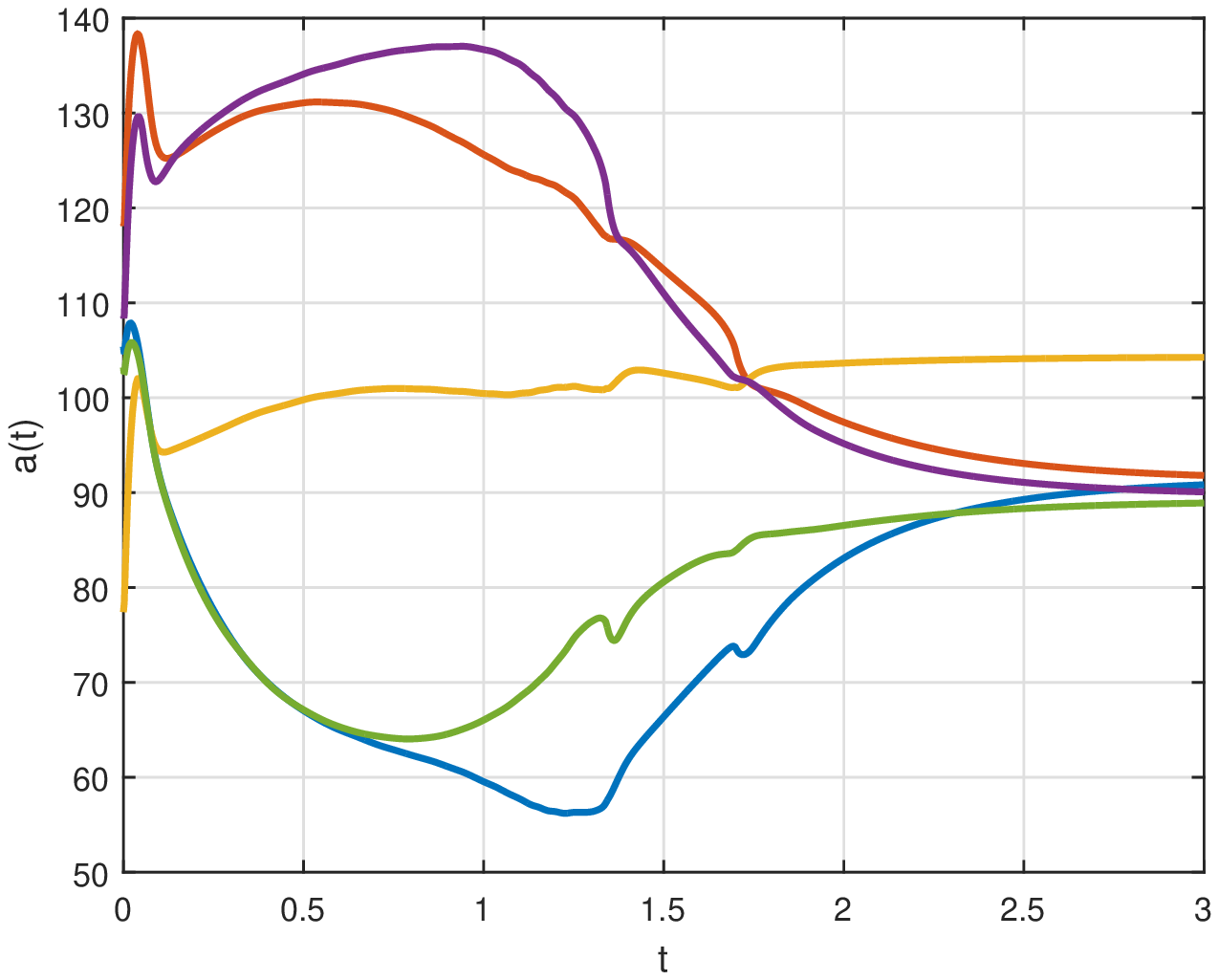}\label{Amp}}
      \subfigure[Phase]{
      \includegraphics[width=.45\textwidth]{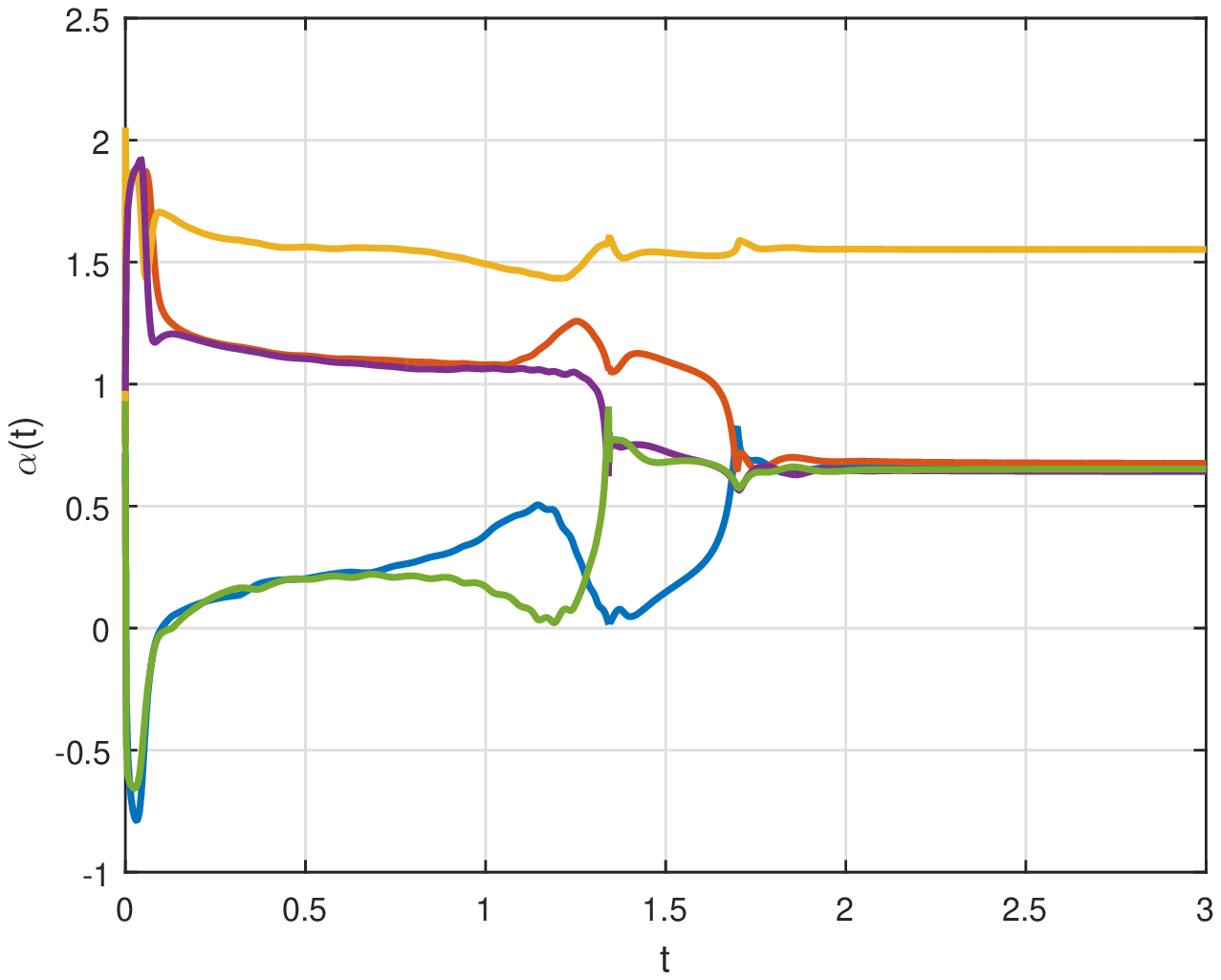}\label{Phs}}
      \caption{Evolution and amplitude and phase all 5 elements during the optimization process}\label{Result2}
  \end{centering}
\end{figure}

Given in Fig.~\ref{Result2} are the parameters obtained from implementing the proposed fast scale optimization algorithm. Fig.~\ref{Amp} contains the amplitude for the each of the 5 array elements while Fig~\ref{Phs} contains the phase.

%%%%%%%%%%%%%%%%%%%%%%%%%%%%%%%%%%%%%%%%%%%%%%%%%%%%%%%%%%%%%%%%%%%%%%%%%%%%
%%%%%%%%%%%%%%%%%%%%%%%%%%%%%%%%%%%%%%%%%%%%%%%%%%%%%%%%%%%%%%%%%%%%%%%%%%%%

%%%%%%%%%%%%%%%%%%%%%%%%%%%%%%%%%%%%%%%%%%%%%%%%%%%%%%%%%%%%%%%%%%%%%%%%%%%%%%%%
\section{Conclusion}\label{sec:conclusion}

Here we presented a two time-scale optimization algorithm for multi-agent coordination for the purpose of distributed beamforming. Each agent is assumed to be randomly positioned with respect to each other with random phase offsets and amplitudes. We propose a two time-scale optimization algorithm that consists of a fast time-scale algorithm to solve for the amplitude and phase while a slow time-scale algorithm to solve for the control required to re-position the agents. The numerical results given here indicate that the proposed two time-scale approach is able to reconstruct a desired beam pattern. Future work include considering positioning as well as synchronization uncertainties and errors. We would also consider the uncertainties associated with channel model and the recruitment problem when the number of required array elements is unknown.
%%%%%%%%%%%%%%%%%%%%%%%%%%%%%%%%%%%%%%%%%%%%%%%%%%%%%%%%%%%%%%%%%%%%%%%%%%%%%%%%

\bibliography{Biblio}
\bibliographystyle{IEEEtran}

\end{document}